\documentclass{ws-ijmpa}
\usepackage{fancybox}
\usepackage{euscript}
\usepackage{amsmath}
\usepackage{amssymb}
\usepackage{color}
\usepackage{cite}

\def\mathswitchr#1{\relax\ifmmode{\mathrm{#1}}\else$\mathrm{#1}$\fi}

\newcommand {\pslash}{\hbox{$\not\hbox{\kern-2.3pt $p$}$}}

\def\rQCED{{\rm QCED}}

\def\alf1{ {\alpha\over\pi} }
\def\rQCD{{\rm QCD}}
\begin{document}

\markboth{\footnotesize B.F.L. WARD, C. GLOSSER, S. JADACH, S. A. YOST}
{THRESHOLD CORRECTIONS IN PRECISION LHC PHYSICS: QED$\otimes$QCD}

%
\catchline{}{}{}{}{}
%

\title{THRESHOLD CORRECTIONS IN PRECISION LHC PHYSICS: QED$\otimes$QCD}

\author{\footnotesize B.F.L. WARD}

\address{Department of Physics, Baylor University, One Bear Place \#97316\\
Waco, Texas 76798-7316,
USA
}

\author{\footnotesize C. GLOSSER}

\address{Department of Physics, Southern Illinois University, Box 1654\\
 Edwardsville, IL 62026-1654,  USA
}
\author{\footnotesize S. JADACH}

\address{Institute of Nuclear Physics, ul. Radzikowskiego 152\\
31-342 Krak\'ow, Poland\\
Theory Division, CERN\\
CH-1211 Geneva 23, Switzerland
}

\author{\footnotesize S. A. YOST}

\address{Department of Physics, Baylor University, One Bear Place \#97316\\
 Waco, Texas 76798-7316, USA}

\maketitle


\begin{abstract}
With an eye toward
LHC processes in which theoretical precisions of 1\% are desired, 
we introduce the 
theory of the simultaneous YFS resummation of QED and QCD
to compute the size of the
expected resummed soft radiative threshold effects in 
precision studies of heavy
particle production at the LHC. Our results 
show that both QED and QCD soft threshold effects
must be controlled to be on 
the conservative side to achieve such precision goals.
\keywords{QCD; QED; Resummation.}
\end{abstract}
\section{Introduction}
At the LHC/ILC,
the precision requirements for soft multiple gluon (n(g)) effects
will be even more demanding than at FNAL, where the uncertainty 
on $m_t$~\cite{mterror}, 
$\delta m_t = 4.3$ GeV, receives a soft n(g) uncertainty $\sim$ 2-3 GeV, 
and soft n(g) MC exponentiation results will be an 
important part of the necessary theory -- YFS exponentiated
${\cal O}(\alpha_s^2)L$ calculations, {\em in the presence of parton showers}, 
on an event-by-event basis.

As many authors~\cite{qcdlit} 
prepare the
necessary results that would lead to such a precision on QCD
for LHC processes, the QED corrections need to be addressed as well.
Estimates by Refs.~\cite{cern2000,spies,james1,roth,james2}
show that one gets few per mille
effects from QED corrections to structure function evolution.
In this paper, estimate the size of QED corrections at threshold
at the LHC.

Treating simultaneously QED and QCD in the respective 
YFS~\cite{yfs,yfs1} exponentiation we discuss
threshold effects at the LHC 
in the candidate luminometry~\cite{lhclum,fnallum} processes
$pp\rightarrow V +n(\gamma)+m(g)+X\rightarrow \bar{\ell} \ell'
+n'(\gamma)+m(g)+X$, where 
$V=W^\pm,Z$,and $\ell = e,\mu,~\ell'=\nu_e,\nu_\mu ( e,\mu )$
respectively for $V=W^+ ( Z )$, and  
$\ell = \nu_e,\nu_\mu,~\ell'= e,\mu$ respectively for $V = W^-$.


\section{YFS Theory and its Extension to QCD}

In Refs.~\cite{yfs1} the renormalization group improved YFS theory~\cite{bflw1yfs} 
for $e^+(p_1)e^-(q_1)\rightarrow \bar{f}(p_2) f(q_2) +n(\gamma)(k_1,\cdot,k_n)$
is realized by Monte Carlo methods, where the respective cross section
$d\sigma_{exp}$ and all of the attendant
IR functions, $\{B,\tilde B, D, \tilde S\}$, and hard photon residuals
,$\{\bar\beta_n(k_1,\ldots,k_n)\}$, are specified in Refs.~\cite{yfs1}.
In Refs.~\cite{qcdref,dglapsyn}
we have extended the YFS theory to QCD:
the net result is that in the analogous YFS theory we have the 
replacements $2\alpha\,Re\,B+2\alpha\,\tilde B\rightarrow SUM_{IR}(QCD)$,
$D\rightarrow D_\rQCD$, and $\bar\beta_n(k_1,\dots,k_n)\rightarrow \tilde{\bar\beta}_n(k_1,\ldots,k_n)$, where the QCD YFS functions are defined
in Ref.~\cite{qcdref} and the
gluon residuals~\cite{qcdref} 
$\tilde{\bar\beta}_n(k_1,\ldots,k_n)$
are free of all infrared divergences to all 
orders in $\alpha_s(Q)$. 
The genuine non-Abelian IR physics is encoded~\cite{qcdref} here
in the $\tilde{\bar\beta}_j$.
The YFS resummation which we discuss here is
fully consistent with that of Refs.~\cite{ster,cattrent}.
See Ref.~\cite{CG1} for more discussion of this point.

\section{Extension to QED$\otimes$QCD and QCED}

Simultaneous exponentiation of
QED and QCD higher order effects
gives~\cite{CG1}
{\small
\begin{equation}
\begin{split}
d\hat\sigma_{\rm exp} &= e^{\rm SUM_{IR}(QCED)}
   \sum_{{m,n}=0}^\infty\int\prod_{j_1=1}^m\frac{d^3k_{j_1}}{k_{j_1}} 
\prod_{j_2=1}^n\frac{d^3{k'}_{j_2}}{{k'}_{j_2}}
\int\frac{d^4y}{(2\pi)^4}\\&e^{iy\cdot(p_1+q_1-p_2-q_2-\sum k_{j_1}-\sum {k'}_{j_2})+
D_\rQCED} 
\tilde{\bar\beta}_{m,n}(k_1,\ldots,k_m;k'_1,\ldots,k'_n)\frac{d^3p_2}{p_2^{\,0}}\frac{d^3q_2}{q_2^{\,0}},
\end{split}
\label{subp15b}
\end{equation}}\noindent
where the new YFS functions, $SUM_{IR}(QCED),~D_\rQCED$ and
$\tilde{\bar\beta}_{m,n}(k_1,\ldots,k_m;k'_1,\ldots,k'_n)$, where the
latter has $m$ hard gluons and $n$ hard photons, are
defined in Ref.~\cite{CG1}.
The infrared algebra QCED~\cite{CG1} obtains: the average Bjorken $x$ values
for the QED and QCD emissions imply~\cite{CG1}
%
that QCD dominant corrections happen an
order of magnitude earlier than those for QED so
that the leading $\tilde{\bar\beta}_{0,0}^{(0,0)}$-level
gives a good estimate of the size of the effects we study.

\section{QED$\otimes$QCD Threshold Corrections at the LHC} 


For the basic formula (we use the standard notation here~\cite{CG1}){\small
\begin{equation}
d\sigma_{exp}(pp\rightarrow V+X\rightarrow \bar\ell \ell'+X') =
\sum_{i,j}\int dx_idx_j F_i(x_i)F_j(x_j)d\hat\sigma_{exp}(x_ix_js),
\label{sigtot} 
\end{equation}}\noindent
we use the result in (\ref{subp15b}) for $V=Z$ here with semi-analytical
methods and structure functions from Ref.~\cite{mrst1}. See also the work of Refs.~\cite{baurall,ditt,russ,van1,van2,anas}.
A Monte Carlo realization will appear elsewhere~\cite{elsewh},
wherein we will ultimately use HERWIG~\cite{herwig}, 
PYTHIA~\cite{pythia} and /or the new shower algorithm in Ref.~\cite{jadskrz} 
in lieu of the $\{F_i\}$ and thereby, in principle, improve on the shower/exact
result combination in Ref.~\cite{frixw}. Due to its lack of the appropriate color coherence~\cite{mmm}, we do not consider ISAJET~\cite{isajet} here.


We compute , with and without QED, the ratio
$r_{exp}=\sigma_{exp}/\sigma_{Born}$
to get the results
(We stress that we {\em do not} use the narrow resonance approximation here.){\small
\begin{equation}
r_{exp}=
\begin{cases}
1.1901&, \text{QCED}\equiv \text{QCD+QED,~~LHC}\\
1.1872&, \text{QCD,~~LHC}\\
1.1911&, \text{QCED}\equiv \text{QCD+QED,~~Tevatron}\\
1.1879&, \text{QCD,~~Tevatron.}\\
\end{cases}
\label{res1}
\end{equation}}\noindent
QED is at the level of .3\% at both LHC and FNAL
\footnote{This is stable under scale variations~\cite{CG1}.}.
We agree with the results in Refs.~\cite{baurall,ditt,russ,van1,van2}.
The QED effect is similar in size to structure function
results in Refs.~\cite{cern2000,spies,james1,roth,james2}.



\section*{Acknowledgments}
Two of us ( S.J. and B.F.L.W.)
thank Profs. S. Bethke and L. Stodolsky for the support and kind
hospitality of the MPI, Munich, while a part of this work was
completed. This work was supported partly by US DoE contract
DE-FG05-91ER40627 and by NATO grants PST.CLG.97751,980342.

\end{document}